\documentclass[english,11pt]{article}
\usepackage[margin=3.0cm]{geometry}
\usepackage{graphicx}
\usepackage{cite}
\usepackage{float}
\usepackage{amsmath}
\usepackage{amssymb}
\usepackage[format=plain,indention=1.2cm,textfont=sl]{caption}
\usepackage[affil-sl]{authblk}

\usepackage{color}

\title{\centering Curvature-based Analysis of Brain Networks}
\author[1,*]{Melanie Weber}
\author[2,3]{Johannes Stelzer}
\author[4,5]{Emil Saucan}
\author[5]{Alexander Naitsat}
\author[2,3]{Gabriele Lohmann}
\author[6,7]{J{\"u}rgen Jost}

\affil[1]{Princeton University, Princeton, NJ, USA}
\affil[2]{Department of Biomedical Magnetic Resonance Imaging, University Hospital T{\"u}bingen, 
 T{\"u}bingen, Germany}
\affil[3]{Magnetic Resonance Centre, Max-Planck-Institute for Biological Cybernetics, 
  T{\"u}bingen, Germany}
\affil[4]{ORT Braude College, Karmiel, Israel}
\affil[5]{Technion - Israel Institute of Technology, Haifa, Israel}
\affil[6]{Max-Planck-Institute for Mathematics in the Sciences, Leipzig, Germany}
\affil[7]{Santa Fe Institute for the Sciences of Complexity, New Mexico, USA}
\affil[*]{Corresponding author: Melanie Weber, email: mw25@math.princeton.edu}

\begin{document}

\date{}

\maketitle 

\begin{abstract}
The human brain forms functional networks on all spatial scales.
Modern fMRI scanners allow to resolve functional brain data in high resolutions,
allowing to study large-scale networks that relate to cognitive processes.
The analysis of such networks forms a cornerstone of experimental neuroscience.
Due to the immense size and complexity of the underlying data sets, efficient evaluation and
visualization remain a challenge for data analysis.
In this study, we combine recent advances in experimental neuroscience and applied mathematics
to perform a mathematical characterization of complex networks constructed
from fMRI data. We use task-related edge densities~\cite{lohmann2016} for constructing networks
of task-related changes in synchronization. This construction captures the dynamic
formation of patterns of neuronal activity and therefore represents efficiently the connectivity structure
between brain regions. 
Using geometric methods that utilize Forman-Ricci curvature as an edge-based network characteristic~\cite{WSJ1},
we perform a mathematical analysis of the resulting complex networks. We motivate the use of edge-based characteristics
to evaluate the network structure with geometric methods. 
The geometric features could aid in understanding the connectivity
and interplay of brain regions in cognitive processes.
\end{abstract}


\section{Introduction}

The key idea of network analysis is to reduce a system to relations among its basic elements
as a means for understanding their interactions and commonalities.
Recent technological advances in the natural sciences allow for an analysis and description of highly complex systems on
ever smaller and more detailed scales. In the neurosciences, novel experimental technologies allow for the analysis
of neural systems on the scale of specific brain regions or even on the single-neuron level.

In the present study we bring together recent advances in Computational Neuroscience and Applied Mathematics to introduce
a new approach in analyzing such large-scale neuro-scientific data. In \cite{lohmann2016}, Lohmann et al. developed a method
for analyzing fMRI data with a network-based approach called ``Task-related Edge Density (TED)''.
The main idea behind TED is to compute large-scale, task-related changes
in brain connectivity on the voxel level. The task-related changes between a pair of voxel is defined as difference
in synchronization between the respective time series and do not depend on any specific haemodynamic response model.
Critically, such changes in synchronization occur in dense packs around the spatial neighborhood of the voxel pair.
Networks resulting from a TED analysis are large, complex objects whose analysis
with traditional tools presents a computational challenge. Here we propose to investigate these networks
using a novel formalism based on the concept of curvature.

This  concept was recently investigated in a series of papers~\cite{WSJ1,WSJ2,SSWJ}, where it was shown that
it provides a new set of characteristics for network analysis. Specifically, Weber et al. 
show that information encoded in the edges yields  insights into the local and global structure and
the direction and density of information flow within a network. These features are not captured in classic node-based
network characteristics, such as the distribution of node degrees (i.e. the number of edges per node) and the average
path length (minimal number of edges connecting any pair of two nodes) across a given network.
The core component of this theory is a discrete Ricci-curvature introduced by R. Forman~\cite{Forman}. Utilizing this Ricci curvature, 
this work suggests network-analytic methods with a wide range of possible applications to data mining.

\section{Methods}

\subsection{Forman-Ricci Curvature for Complex Networks}

Traditionally, network analysis focuses on the extraction features of nodes such as 
node degree or the local clustering coefficient. 
Here we propose a novel approach that instead extracts features of edges rather than of nodes. Specifically,
we make use of the notion of {\em curvature}.  Various
notions of discrete curvature have been explored on networks, starting with the combinatorial analogue
of the classical Gauss curvature, in the guise of the well known clustering coefficient.

Among the various notions of curvature, a particularly powerful as well as flexible type of curvature has emerged,
namely that of Ricci curvature. While it does not represent one of the most commonly known versions of curvature,
it has proven itself lately to be a tool that is strong enough
to capture deep phenomena in networks, while at the same time being simple and adaptive enough to make it useful in various discrete settings.
Two different discretizations of Ricci curvature, have been shown to be very efficient in the geometrization of networks, thence effective
in their analysis and even in the prediction of their long time behavior. The better know of these is Ollivier's (coarse)
Ricci curvature~\cite{Ol1,Ol2},
which has seen various applications in many practical fields~\cite{Allen2,GGL}. The second one, based on Forman's theoretical work on the
so called weighted CW complexes, was proposed recently by some of the authors~\cite{SMJSS,WSJ1}, and it has been shown to be easy to define,
simple to manipulate, yet, at the same time, quite powerful.
It is defined as follows.

\medskip

\noindent
We denote a network graph by $G=\lbrace V, E, \omega \rbrace$ where $V :=\lbrace v \in G \rbrace$ is the set of nodes (or vertices),
$E := \lbrace e=(v_1 , v_2) \; \forall v_1, v_2 \in G\rbrace$ the set of edges each connecting two nodes.
Let further $\omega := \lbrace \omega (V), \omega (E)\rbrace$ be the weighting schemes on nodes ($\omega (V)$) and edges ($\omega (E)$),
both with values in $[0,1]$. Then the Forman-Ricci curvature is defined by

\begin{align}
 {\rm Ric}_F (e) = \omega(e) \left( \frac{\omega(v_1)}{\omega(e)} + \frac{\omega(v_2)}{\omega(e)}
 - \sum_{\substack{\omega(e_{v_1}) \sim e \\ \omega(e_{v_2}) \sim e}} \left( \frac{\omega(v_1)}{\sqrt{\omega(e)\omega(e_{v_1})}}
 + \frac{\omega(v_2)}{\sqrt{\omega(e)\omega(e_{v_2})}} \right) \right)
 \label{eq:FR-curv}
 \end{align} 

for an edge $e=(v_1, v_2) \in E$ and $e_{v_1}, e_{v_2} \in E$ any edges adjacent to $e$ at vertices $v_1$ and $v_2$.
For unweighted graphs, this simplifies to
\begin{align}
{\rm Ric}_F (e) = 4 - d_{i} - d_{j}
 \label{eq:FR_unweighted}
\end{align} 
\noindent where  $i,j$ are the two vertices connected by the edge $e$, and  $d_{i}, d_{j}$ denote their degrees.
Here we assume that all vertex and edge weights are set to~1. The following figure presents the geometric intuition behind this concept. 

\begin{figure*}[h!]
\begin{center}
\includegraphics[width=.5\columnwidth]{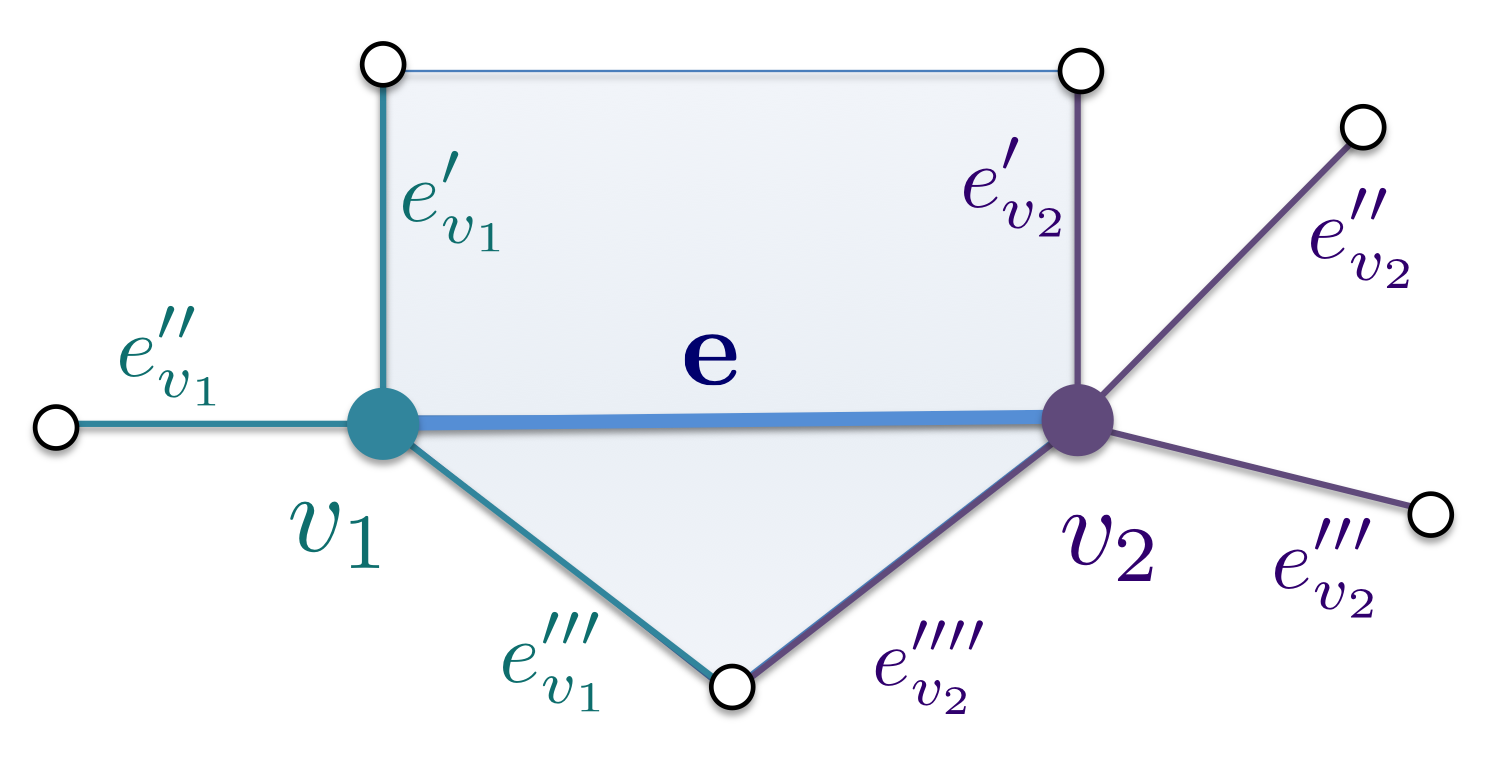} 
	\caption{{\bf Local connectivity structure.} Edge $e$ with adjacent vertices $v_1$ and $v_2$
        and parallel edges $\lbrace e_{v_1}', e_{v_1}'', e_{v_1}''' \rbrace$ (adjacent to $v_1$)
        and $\lbrace e_{v_2}', e_{v_2}'', e_{v_2}''', e_{v_2}'''' \rbrace$ (adjacent to $v_2$).}
\end{center}
	\label{fig:FR-sketch}
\end{figure*}

In unweighted graphs, curvature values are in the range $(-\infty,4]$, and 
edges connecting vertices of large degree have very negative curvature values.
Such edges may be interpreted as being most important for the cohesion of the network and form
the ``backbone'' of the graph. In the experiments reported below, we only use unweighted graphs.

Besides looking at the most curved edges, we also investigate the statistics of the distributions of curvature values, and this will lead us to the identification of particular subnetworks.

\subsection{Task-related edge density (TED)}

\begin{figure*}[h!]
\centerline{
\includegraphics[width=0.4\textwidth]{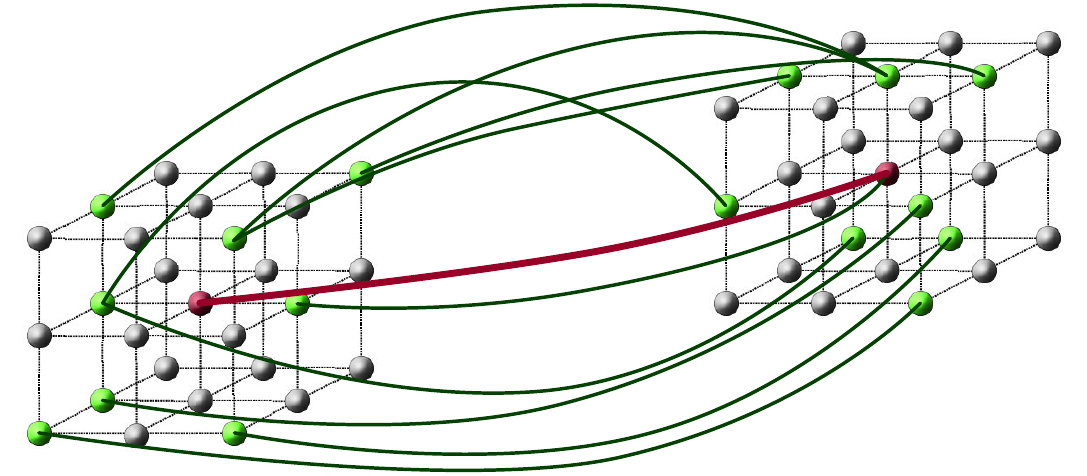} \hspace*{0.5cm}
\includegraphics[width=0.41\textwidth]{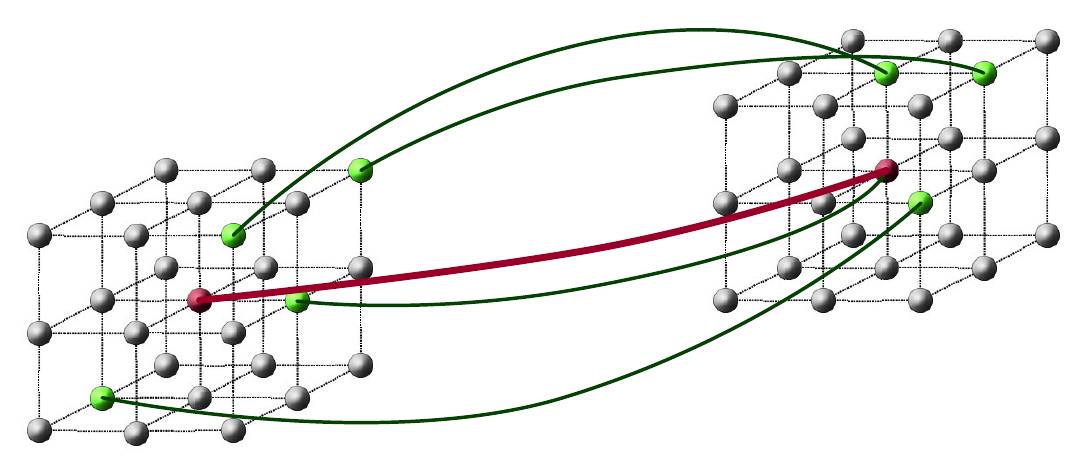}}  
\caption{{\bf Illustration of task-related edge densities (from~\cite{lohmann2016}).}
An edge has a high density if the percentage of suprathreshold edges in its local neighbourhood is high. 
Thus, the red edge shown on the left has a high density, whereas the red edge on the right has a low density.
The local neighbourhood of an edge is defined as the set of edges that connect the 18-adjacent voxels of its endpoints.
The TED algorithm computes edge densities for all suprathreshold edges. In our experiments, the threshold was set
to the top 1~percent of all functional correlations.}
\label{ted}
\end{figure*}

Task-related edge density (TED)~\cite{lohmann2016} is a novel way for investigating changes in
functional connectivity across a set of experimental conditions. It allows a whole-brain investigation
into changes of connectivity and thus, it is not necessary to define a seeding region. Furthermore, the
method operates on the voxel level, rendering pre-segmentations obsolete. TED relies on changes in
synchronization between pairs of voxels and does not make assumptions on the haemodynamic response
function.

The key idea behind TED is the observation that if two voxels change their synchronization, their
spatial neighborhood also changes their connectivity to a much greater extent than it would be expected
given the inherent spatial correlation between spatially adjacent voxels. Figure~\ref{ted} illustrates this
effect, showing neighborhoods of voxels and their connectivity. The left pair of neighborhoods is
more strongly connected than the one on the right. This reflects on the TED value, which is defined as the
number of connections divided by the number of theoretically possible connections.
Here, the number of possible connections is~729. Thus, the left edge
has a density of 11/729$\approx$0.0151, and the edge on the right has a density of 5/729$\approx$0.007.

\section{Experimental results}

We analyzed task-based fMRI data of provided by the Human Connectome Project (HCP),  WU-Minn Consortium~\cite{HCP,Barch2013}
using minimally preprocessed data of 400 participants of the ``emotion task''.  For details regarding this task see~\cite{Barch2013}.
All data sets were acquired with the following parameters: TR=720ms, TE=33.1ms, 2mm isotropic voxel size, multiband factor~8.
The preprocessing protocol is described in~\cite{HCPpreproc}. To reduce dimensionality, we further downsampled the data to $(3mm)^3$.

Out of the cohort of 400 subjects, we randomly selected ten samples of 100~subjects each (with replacement).
This allowed us to assess the reproducibility of the results.
For each of these ten samples, we computed a network of edges that respond differentially to the ``faces minus shapes'' contrast
of the emotion task resulting in 10~separate networks.
As described above, we used the task-related edge density method (TED)~\cite{lohmann2016} for this purpose.
The initial threshold was set to 0.99, i.e. the top 1~percent of all correlation values were used. The edge density
threshold was set to 0.25, i.e. only edges in which the edge density exceeded 0.25 were included in the resulting
TED networks.

\begin{figure*}[h!]
\centering
\includegraphics[width=0.65\textwidth]{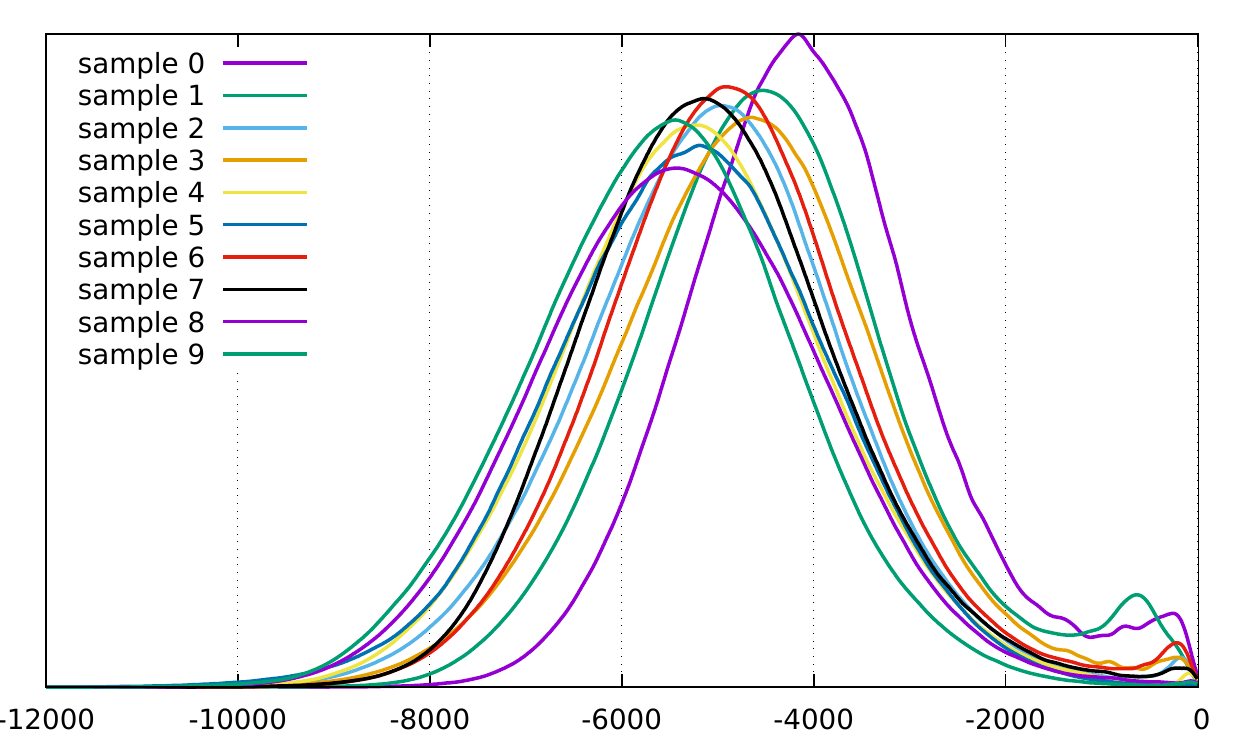}
\caption{{\bf Distributions of Forman-Ricci curvature values in 10~networks.}}
\label{hist}
\end{figure*}

The TED networks were then subjected to the Forman-Ricci curvature analysis as described above.
This resulted in curvature values for all edges of all ten networks.
Fig.~\ref{hist} shows their distributions. They all showed a broad-scale distribution with 
a smaller secondary peak with curvature values less than $-500$. Thus, the distributions deviated clearly from the
classic power law distribution that characterizes real-world networks with a small-world property.

In order to further investigate the tail ends of the curvature distributions, we extracted edges with curvature values $< -4000$,
and also with curvature values $> -500$.
To visualize the results, we projected the selected edges onto ``hubness maps''. A voxel in a hubness map
records the number of edges for which this voxel serves as an endpoint.
Voxels in which many edges accumulate may be viewed as hubs~\cite{lohmann2016}.

\begin{figure*}[h!]
\begin{center}
\includegraphics[width=.7\columnwidth]{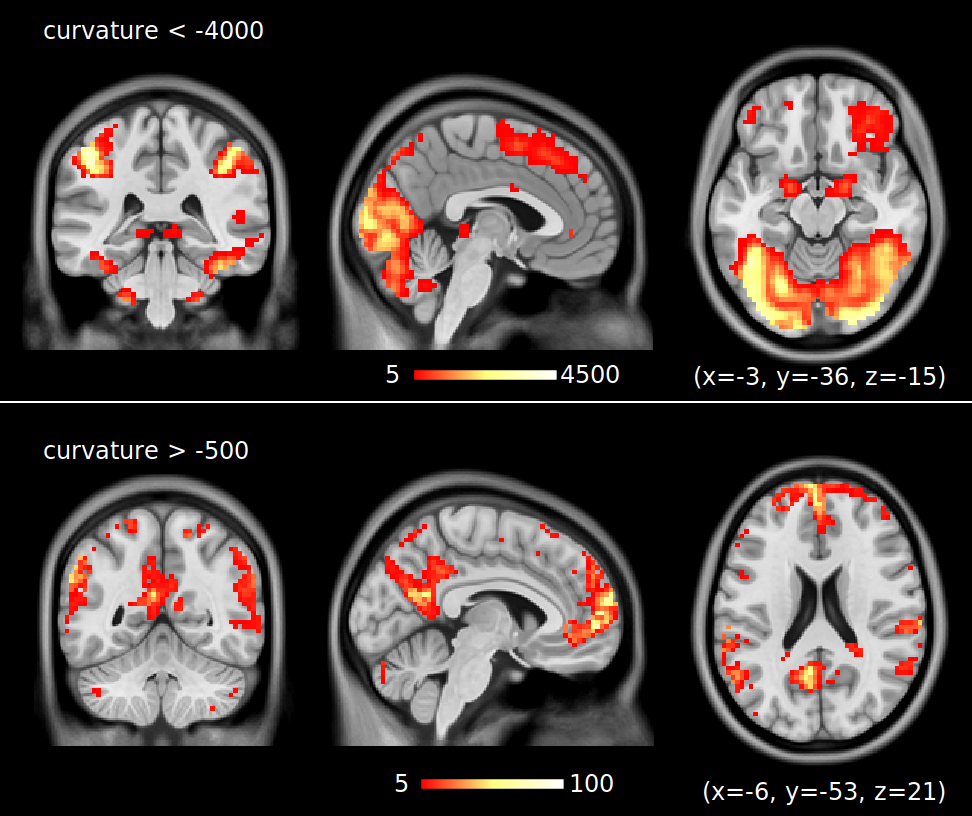}
\end{center}
\caption{{\bf Average hubness maps of very low and very high curvature values.}
The top image shows an average across the 10~hubness maps using edges with curvatures $<-4000$
corresponding to the lower end of the distribution. This image approximately shows
the core the ``face-shape'' contrast of the emotion task.
The lower image shows the upper end of the distribution with curvature values $>-400$.
It appears to show the default mode network.
The images are sliced at MNI coordinates x=-3,y=-36,z=-10 (top) and x=-6,y=-53,z=21 (bottom) }
\label{hubs}
\end{figure*}

Averages across the hubness maps of all 10~networks are shown in Fig.~\ref{hubs}.
At one end of the curvature distribution ($> -500$), the hubness maps appear to show the default mode network~\cite{DMN}.
At the other end ($< -4000$), the  hubness maps show the core of the emotion contrast.
Finally, we investigated the correlation between the edge density measure and the Forman-Ricci curvature.
We found an overall strong negative correlation between both measures when all edges were included.
Specifically, we found an average correlation of $\mu=-0.565, \sigma=0.020$  (averaged across all 10~networks).
For curvature values $< -4000$, the average correlation was $\mu=-0.477, \sigma=0.0489$,
However, this correlation was much weaker around the secondary peak of the distribution (curvature values $> -500$).
Here we found $\mu=-0.164, \sigma=0.097$.

\section{Discussion}

We applied a novel network metric based on the Forman-Ricci curvature concept to networks detected
by the task-related edge density approach (TED) in fMRI data of the human brain.
In contrast to traditional network metrics, this new approach attaches a score to edges rather
than to nodes. It thus has conceptual similarities with the edge clustering coefficient $ECC$
recently proposed in~\cite{Wang2012}. The advantage of the Forman-Ricci curvature concept
is that it has a well established mathematical foundation.

When investigating the distributions of the curvature values, we found that
at curvature values $> -500$ (the secondary peak) the correlation
between edge density and curvature was very weak indicating that the two measures represent different types of
information. Furthermore, we found that those edges appear to
belong to the default mode network (DMN). This is a surprising result
because so far the DMN has been primarily observed in the context of resting state fMRI
whereas here it appears in a task contrast.
However, further experiments are needed to investigate this effect in more detail.

\bigskip

\subsection*{Acknowledgments}
This work was partly funded by the European Commission H2020-GA-634541 CDS-QUAMRI.\\
Data were provided by the Human Connectome Project, WU-Minn Consortium (Principal Investigators:
David Van Essen and Kamil Ugurbil; 1U54MH091657)
funded by the 16 NIH  Institutes and Centers that support the NIH Blueprint for Neuroscience Research;
and by the McDonnell Center for Systems Neuroscience at Washington University.


\bibliography{lit}

\end{document}